\documentclass[twocolumn]{aastex631}

\begin{document}

\title{Small-amplitude Compressible Magnetohydrodynamic Turbulence Modulated by Collisionless Damping in Earth’s Magnetosheath: Observation Matches Theory}

\author[0000-0003-4268-7763]{Siqi Zhao}
\affiliation{Deutsches Elektronen Synchrotron DESY, Platanenallee 6, D-15738, Zeuthen, Germany}
\affiliation{Institut für Physik und Astronomie, Universität Potsdam, D-14476, Potsdam, Germany}

\author[0000-0003-2560-8066]{Huirong Yan}
\affiliation{Deutsches Elektronen Synchrotron DESY, Platanenallee 6, D-15738, Zeuthen, Germany}
\affiliation{Institut für Physik und Astronomie, Universität Potsdam, D-14476, Potsdam, Germany}

\author[0000-0003-1778-4289]{Terry Z. Liu}
\affiliation{Department of Earth, Planetary, and Space Sciences, University of California, Los Angeles, CA 90024, USA}
 \email{huirong.yan@desy.de; terryliuzixu@ucla.edu}

\author[0000-0003-1683-9153]{Ka Ho Yuen}
\affiliation{Theoretical Division, Los Alamos National Laboratory, Los Alamos, NM 87545, USA}

\author[0000-0002-9201-5896]{Mijie Shi}
\affiliation{Shandong Key Laboratory of Optical Astronomy and Solar-Terrestrial Environment, Institute of Space Sciences, 264209, Shandong University, Weihai, People’s Republic of China}



\begin{abstract}
Plasma turbulence is a ubiquitous dynamical process that transfers energy across many spatial and temporal scales and affects energetic particle transport. Recent advances in the understanding of compressible magnetohydrodynamic (MHD) turbulence demonstrate the important role of damping in shaping energy distributions on small scales, yet its observational evidence is still lacking. This study provides the first observational evidence of substantial collisionless damping (CD) modulation on small-amplitude compressible MHD turbulence cascade in Earth's magnetosheath using four \textit{Cluster} spacecraft. Based on an improved compressible MHD decomposition algorithm, turbulence is decomposed into three eigenmodes: incompressible Alfvén modes, and compressible slow and fast (magnetosonic) modes. Our observations demonstrate that CD enhances the anisotropy of compressible MHD modes because CD has a strong dependence on wave propagation angle. The wavenumber distributions of slow modes are mainly stretched perpendicular to the background magnetic field ($\mathbf{B_0}$) and weakly modulated by CD. In contrast, fast modes are subjected to a more significant CD modulation. Fast modes exhibit a weak, scale-independent anisotropy above the CD truncation scale. Below the CD truncation scale, the anisotropy of fast modes enhances as wavenumbers increase. As a result, fast mode fractions in the total energy of compressible modes decrease with the increase of perpendicular wavenumber (to $\mathbf{B_0}$) or wave propagation angle. Our findings reveal how the turbulence cascade is shaped by CD and its consequences to anisotropies in the space environment.

\end{abstract}


\section{Introduction} 

Plasma turbulence, particularly its compressible component, plays a crucial role in numerous astrophysical processes, such as the heating and acceleration of solar wind, cosmic ray transport, and star formation \citep{Bruno2013,Yan2021}. The current model of plasma turbulence is typically characterized by three steps: (1) energy injection on large scales \citep{Matthaeus1986,Cho2003}, (2) inertial energy cascade following some self-similar power law scaling \citep{Ng1996,Horbury2008}, and (3) dissipation caused by certain kinetic physical processes on small scales \citep{Leamon1998,Leamon1999,Yan2002,Alexandrova2009,Sahraoui2009}. Inertial energy cascade, the most characteristic signature of magnetohydrodynamic (MHD) turbulence, has been effectively described using incompressible MHD models such as the isotropic theory (IK65) \citep{Iroshnikov,Kraichnan1965} and scale-dependent anisotropic turbulence theory (GS95) \citep{Goldreich1995}. The nearly incompressible (NI) theory has also been used to explain some phenomena related to compressible solar wind turbulence \citep{Zank1992,Zank1993}. However, within the inertial energy cascade, compressible MHD turbulence undergoes damping processes \citep{Barnes1966,Barnes1967,Yan2004,Suzuki2006}, which is still not completely understood. Fully comprehending how damping affects compressible MHD turbulence is integral for portraying turbulence in actual plasma environment.


One prominent feature of MHD turbulence is the anisotropy, which has been extensively studied through simulations and satellite observations \citep{Cho2000,Stawarz2009,Oughton2015,Huang2017,Andres2022,Jiang2023}. In a homogeneous plasma with a uniform background magnetic field ($\mathbf{B_0}$), small-amplitude compressible MHD fluctuations can be decomposed into three linear eigenmodes (namely, Alfvén mode, slow magnetosonic mode, and fast magnetosonic mode) \citep{Glassmeier1995,Cho2003,Chaston2020,Makwana2020,Zhu2020,Zhao2021,Zhao2022,Zhao2023}. The linear independence among the three MHD eigenmodes enables individual analysis of their statistical properties of small amplitude plasma turbulence \citep{Cho2003,Cho2005}. The composition of MHD modes significantly affects the energy cascade and observational turbulence statistics \citep{Andres2018,Makwana2020,Zhang2020,Brodiano2021,Zhao2022,Malik2023,Yuen2023,4DFFT_p1}. Based on the modern theory of compressible MHD turbulence, Alfvén and slow modes are expected to follow a cascade with scale-dependent anisotropy $k_{\parallel} \propto k_{\perp}^{2/3}$, where $k_{\perp}$ and $k_{\parallel}$ are wavenumbers perpendicular and parallel to $\mathbf{B_0}$ \citep{Goldreich1995,Lithwick2001}. In contrast, fast modes are expected to show isotropic behaviors and cascade like the acoustic wave \citep{Cho2003,Galtier2023}. These theoretical conjectures have been confirmed by numerical simulations \citep{Cho2005,Makwana2020}.

Earlier theoretical studies have demonstrated a strong propagation angle dependence on collisionless damping and viscous damping, influencing the three-dimensional (3D) energy distributions \citep{Yan2004,YLD2004,Petrosian2006,Yan2008}. The collisionless damping (CD) leads to the rapid dissipation of plasma waves by wave-particle interactions via gyroresonance, transit time damping, or Landau resonance \citep{Yan2004,Yan2008}. Despite theoretical predictions, direct observations demonstrating how CD modulates the statistics of compressible MHD modes are still lacking, primarily due to the limited satellite measurements. Thanks to the availability of spatial information from four \textit{Cluster} spacecraft, we can estimate energy distributions of compressible turbulence. Although early Cluster data are relatively old, we use a novel, improved MHD mode decomposition method to analyze them. This study compares the theoretical CD rate and observed energy distributions, providing the first observational evidence of substantial CD modulation on small-amplitude compressible MHD turbulence.

\begin{figure}
\includegraphics[scale=0.4]{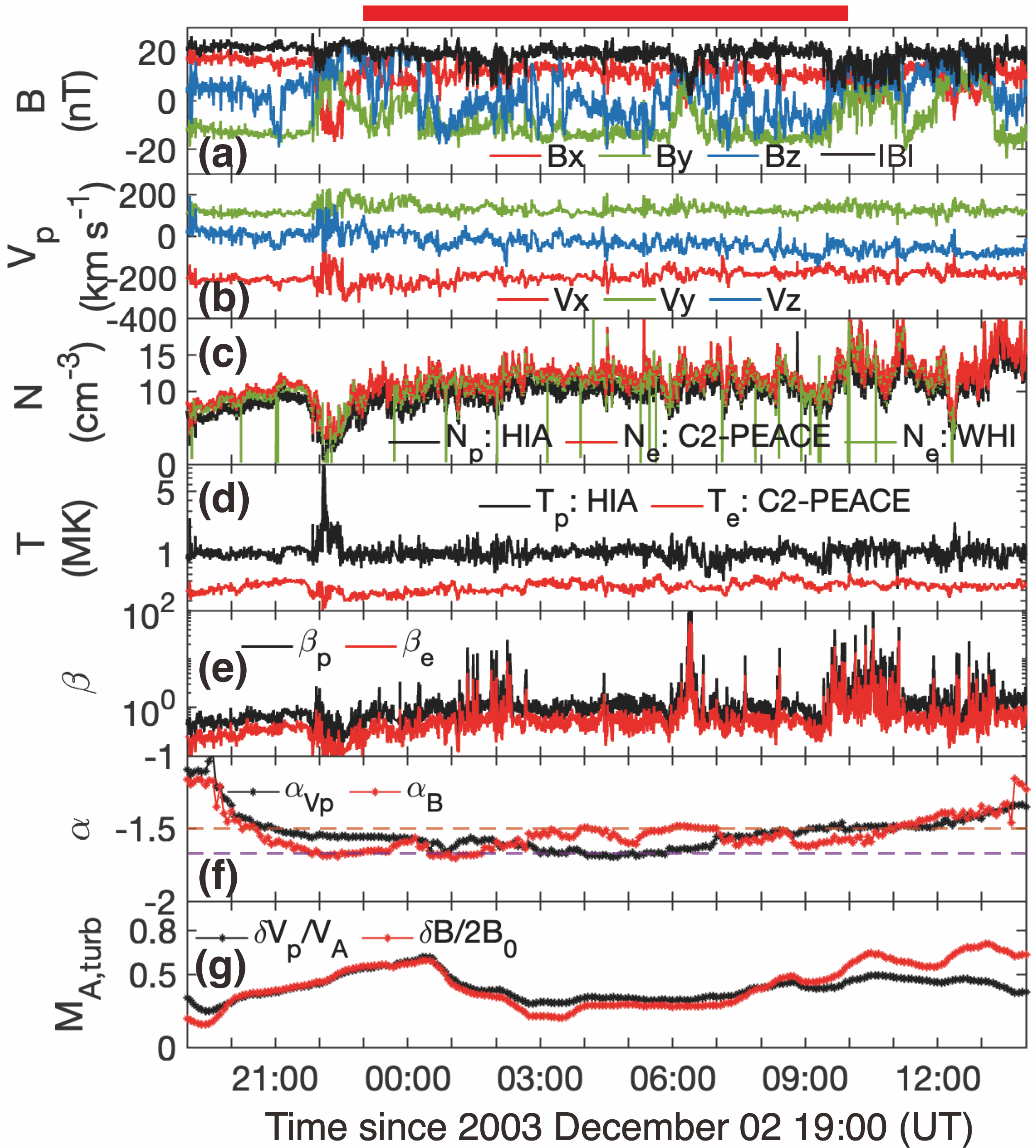}
\caption{An overview of fluctuations measured by \textit{Cluster}-1. The red bar on the top marks the interval during 23:00-10:00 UT on 2-3 December 2003. (a) Magnetic field. (b) Proton bulk velocity. (c) Cross-check of proton density from CIS-HIA (black), and electron density from Plasma Electron And Current Experiment (PEACE) \citep{Johnstone1997} onboard \textit{Cluster}-2 (red) and from Waves of High frequency and Sounder for Probing of Electron density by Relaxation (WHISPER) \citep{Decreau1997} (green). (d,e) Temperature and plasma $\beta$ (the ratio between thermal and magnetic pressures). (f) Spectral slopes ($\alpha$) of trace proton velocity and magnetic field power. The horizontal dashed lines represent $\alpha=-5/3$ and $-3/2$. (g) $\delta V_p/V_A$ (black) and $\delta B/(2B_0)$ (red), where $\delta V_p$ and $\delta B$ are root mean squared (rms) proton velocity and magnetic field fluctuations, respectively.}
\end{figure}

\section{Overview} 
Figure 1 shows an overview of \textit{Cluster} observations in Earth’s magnetosheath during 19:00-14:00 UT on 2-3 December 2003 in Geocentric Solar Ecliptic (GSE) coordinates. During this time interval, the \textit{Cluster} mission is in a tetrahedral-like configuration, with the relative separation $d_{sc}\sim200\rm{km}$ (around 3 proton inertial length $d_i\sim74\rm{km}$), enabling us to perform a multi-point analysis on MHD turbulence. To ensure low-frequency (large-scale) measurements while the mean magnetic field $\mathbf{B}_0$ approaches the local background magnetic field, we split the whole time interval into several time windows with a five-hour length and a five-minute moving step. Performing an MHD mode decomposition during 23:00-10:00 UT (the red bar on the top of Figure 1) is applicable for the following reasons. First of all, the background magnetic field measured by the Fluxgate Magnetometer (FGM) \citep{Balogh1997} and proton plasma parameters measured by the Cluster Ion Spectrometry’s Hot Ion Analyzer (CIS-HIA) \citep{Reme2001} are relatively stable, as shown in Figures 1(a-d). Fluctuations are approximately stationary and homogeneous based on the analysis of correlation functions (see Appendix A). Additionally, fluctuations are in a well-developed state, as shown in Figure 1(f) where the spectral slopes ($\alpha_{V_p}$ and $\alpha_B$) of the trace proton velocity and magnetic field power at spacecraft-frequency $f_{sc}\sim[0.001 Hz,0.1 f_{ci}$] range between $-5/3$ and $-3/2$ (the proton gyro-frequency $f_{ci}\sim0.24 Hz$). The trace proton velocity and magnetic field power are calculated through the fast Fourier transform (FFT) with five-point smoothing in each time window. Figure 1(g) shows that the turbulent Alfvén Mach number ($M_{A,turb}=\delta V_p/V_A$) and relative amplitudes of magnetic field fluctuations ($\delta B/B_0$) are smaller than unity, indicating that the nonlinear terms ($\delta \mathbf{V}_p^2$, $\delta \mathbf{B}^2$) are smaller than the linear terms ($V_A\delta \mathbf{V}_p$,$B_0\delta \mathbf{B}$). Thus, the assumption of small-amplitude approximation is satisfied. Table \ref{tab:table1} lists the values of background physical parameters for our analysis.

\begin{table*}
\caption{\label{tab:table1} Physical parameters for this observed event. Outliers, defined as elements more than three scaled median absolute deviations ($1.48\cdot \text{median}(|X-\text{median}(X)|)$) from the median, are replaced with the linear interpolation of neighboring, non-outlier values.}
\begin{ruledtabular}
\begin{tabular}{cccccccccc}
 Start time (UT)&End time (UT)&$B_0$ (nT)&$N_0$ $(cm^{-3})$
&$V_A$ $(km/s)$&$V_S$ $(km/s)$&$\beta_p$& $V_{Te}$ $(km/s)$&$d_i$ $(km)$&$f_{ci}$ $(Hz)$\\ \hline
 2003-12-02/23:00&2003-12-03/10:00&19.2&10.1&133&121&1.0&2567&74&0.24 \\
\end{tabular}
\end{ruledtabular}
\end{table*}

\section{MHD mode decomposition} 

Previous studies suggest that strong turbulence develops scale-dependent anisotropy in the local frame of reference \citep{Cho2000}. In order to trace the local frame anisotropy, we split the whole time interval into several time windows. In each time window, we separate compressible fluctuations into slow and fast modes (Alfvén modes are analyzed in \citealt{Zhao2023}) and establish 3D wavenumber distributions. We decompose MHD eigenmodes by combining three methods: linear decomposition method \citep{Cho2003,Zhao2021}, singular value decomposition (SVD) method \citep{Santolik2003}, and multi-spacecraft timing analysis \citep{Grinsted2004}. The combination of the three methods allows direct retrieval of energy wavenumber distributions from the observed frequency distributions independent of any spatiotemporal hypothesis (e.g., Taylor hypothesis, \citealt{Taylor1938}).  

First, we obtain wavelet coefficients of proton velocity, magnetic field, and proton density using Morlet-wavelet transforms \citep{Grinsted2004}. To eliminate the edge effect resulting from finite-length time series, we perform wavelet transforms twice the length of the studied period and remove the affected periods.

Second, wavevector directions ($\hat{\mathbf{k}}_{SVD} (t,f_{sc})$) are calculated by creating a matrix equation ($\mathbf{A}\cdot \hat{\mathbf{k}}_{SVD}=0$) equivalent to the linearized Gauss’s law ($\mathbf{B} \cdot \hat{\mathbf{k}}_{SVD}=0$). The matrix $\mathbf{A}$ ($6\times3$) consists of the real and imaginary parts of the spectral matrix $S_{mn}$, where $S_{mn}=B_mB_n^*$, $m=X,Y,Z$ and $n=X,Y,Z$ are three components in GSE coordinates (for details, see Equation (8) in \citealt{Santolik2003}). The energy density at each time $t$ and $f_{sc}$ is a mixture of fluctuations with different dispersion relations (and thus different wavevectors). However, the SVD method only gives the best-estimated direction of the wavevector sum but not the wavevector magnitude. The unit wavevectors calculated by SVD and mean magnetic field are averaged over four \textit{Cluster} spacecraft: $\mathbf{k}_{SVD}=\frac{1}{4}\sum_{i=1,2,3,4}\hat{\mathbf{k}}_{SVD,Ci}$ and $\mathbf{B}_0=\frac{1}{4}\sum_{i=1,2,3,4}\mathbf{B}_{0,Ci}$, where $Ci$ denotes the four \textit{Cluster} spacecraft.


Third, the $\hat{k}\hat{b}_0$ coordinates are determined by $\hat{\mathbf{k}}_{SVD}=\mathbf{k}_{SVD}/|\mathbf{k}_{SVD}|$ and $\hat{\mathbf{b}}_0=\mathbf{B}_0/|\mathbf{B}_0|$ (see Figure 6(a) in Appendix). The axis basis vectors are $\hat{\mathbf{e}}_{\parallel}=\hat{\mathbf{b}}_0$
, $\hat{\mathbf{e}}_{\perp1}=\hat{\mathbf{k}}_{SVD}\times \hat{\mathbf{b}}_0/|\hat{\mathbf{k}}_{SVD}\times \hat{\mathbf{b}}_0|$, and $\hat{\mathbf{e}}_{\perp2}=\hat{\mathbf{b}}_0\times(\hat{\mathbf{k}} _{SVD}\times\hat{\mathbf{b}}_0)/|\hat{\mathbf{b}}_0\times(\hat{\mathbf{k}}_{SVD}\times\hat{\mathbf{b}}_0)|$. The complex vectors (wavelet coefficients of the proton velocity and magnetic field) are transformed from GSE coordinates to the $\hat{k}\hat{b}_0$ coordinates. 

Fourth, magnetic field data are measured by four \textit{Cluster} spacecraft, whereas proton plasma data are only available on \textit{Cluster}-1 during the analyzed period. Thus, magnetic field power is calculated by $P_{B_l}(t,f_{sc})=\frac{1}{4}\sum_{i=1,2,3,4}W_{B_l,Ci}W_{B_l,Ci}^*$, where $l$ represents $\hat{\mathbf{e}}_{\parallel}$, $\hat{\mathbf{e}}_{\perp1}$, and $\hat{\mathbf{e}}_{\perp2}$. The proton velocity and proton density power are calculated by $P_{V_l}(t,f_{sc})=W_{V_l,C1}W_{V_l,C1}^*$ and $P_{N}(t,f_{sc})=W_{N,C1}W_{N,C1}^*$. $W_{V_l}$, $W_{B_l}$, and $W_{N}$ represent wavelet coefficients of proton velocity, magnetic field, and proton density fluctuations. 

Fifth, noticing that SVD does not give the magnitude of wavevectors, we utilize multi-spacecraft timing analysis based on phase differences between magnetic wavelet coefficients to determine wavevectors ($\mathbf{k}_{B_l}(t,f_{sc})$) \citep{Pincon2008}. The magnetic field data are interpolated to a uniform time resolution of $2^3 samples/s$ for sufficient time resolutions. We consider that the wave front is moving in the direction $\mathbf{\hat{n}}$ with velocity $V_{w}$. The wavevectors $\mathbf{k}_{B_l}=2\pi f_{sc}\mathbf{m}$ are determined by phase differences of magnetic field $B_l$ component,
\begin{eqnarray}
  \left(
  \begin{array}{cc}
      \mathbf{r}_{2} - \mathbf{r}_{1} \\
      \mathbf{r}_{3} - \mathbf{r}_{1} \\
      \mathbf{r}_{4} - \mathbf{r}_{1}
  \end{array} 
  \right) \mathbf{m} = 
  \left(
 \begin{array}{cc}
      \delta t_2 \\
      \delta t_3 \\
      \delta t_4 
 \end{array} 
  \right)
  \label{eq:1}
\end{eqnarray}
where the vector $\mathbf{m}=\mathbf{\hat{n}}/{V_{w}}$, and \textit{Cluster}-1 has arbitrarily been taken as the reference  \citep{Pincon2008}. The left side of Eq.(\ref{eq:1}) is the relative spacecraft separations. The right side of Eq.(\ref{eq:1}) represents the weighted average time delays, estimated by the ratio of six phase differences ($\phi_{ij}=arctan(\mathcal{S}(W_{B_l}^{ij}),\mathcal{R}(W_{B_l}^{ij}))$) to the angular frequencies ($\omega_{sc}=2\pi f_{sc}$), where $\phi_{ij}$ is from all spacecraft pairs ($ij=12$, $13$, $14$, $23$, $24$, $34$)). $\mathcal{S}$ and $\mathcal{R}$ are the imaginary and real parts of cross-correlation coefficients, respectively. Four \textit{Cluster} spacecraft provide six cross-correlation coefficients for each $B_l$ component \citep{Grinsted2004}, i.e., $W_{B_l}^{12}=\langle W_{B_l,C1}W_{B_l,C2}^*\rangle$, $W_{B_l}^{13}=\langle W_{B_l,C1}W_{B_l,C3}^*\rangle$, $W_{B_l}^{14}=\langle W_{B_l,C1}W_{B_l,C4}^*\rangle$, $W_{B_l}^{23}=\langle W_{B_l,C2}W_{B_l,C3}^*\rangle$, $W_{B_l}^{24}=\langle W_{B_l,C2}W_{B_l,C4}^*\rangle$, and $W_{B_l}^{34}=\langle W_{B_l,C3}W_{B_l,C4}^* \rangle$, where $\langle...\rangle$ denotes a time average over $256s$ for the reliability of phase differences.

It is worth noting that multi-spacecraft timing analysis determines actual wavevectors of each $B_l$ component ($\mathbf{k}_{B_l}=\mathbf{k}_{B_\parallel}$, $\mathbf{k}_{B_{\perp1}}$, and $\mathbf{k}_{B_{\perp2}}$), different from the best-estimated direction of the sum of wavevectors determined by three magnetic field components ($\hat{\mathbf{k}}_{SVD}$). Therefore, $\mathbf{k}_{B_l}$ is not completely aligned with $\mathbf{\hat{k}}_{SVD}$. During the analyzed period, Alfven-mode fluctuations dominate ($\delta V_{\perp1}/\delta V_p \sim 60\%$); thus, the $\hat{k}\hat{b}_0$ coordinates determined by $\hat{\mathbf{k}}_{SVD}$ are little affected only when $\mathbf{k}_{B_\parallel}$ and $\mathbf{k}_{B_{\perp2}}$ coplanar with the $\hat{k}\hat{b}_0$ plane, respectively. In other words, $\mathbf{k}_{B_\parallel}$ and $\mathbf{k}_{B_{\perp2}}$ should deviate from the $\hat{k}\hat{b}_0$ plane by a small angle, respectively. To simplify, We shall write $\mathbf{k}$ as a shorthand notation for $\mathbf{k}_{B_l}$ and define the angle between $\mathbf{k}$ and the $\hat{k}\hat{b}_0$ plane as $\eta$. This study only analyzes fluctuations with small $\eta$. With a more stringent $\eta$ constraint, the mode decomposition between slow and fast modes becomes more complete, whereas the uncertainties caused by limited samplings increase. We discuss the uncertainties resulting from the $\eta$ deviation in Section 6 (see color-shaded regions in Figure 5) and Appendix B, and present spectral results under $\eta<20^\circ$ in the main text. Overall, the main properties of energy spectra of slow and fast modes do not significantly change as $\eta$ varies. 

Sixth, we construct a set of $200\times200\times200$ bins to obtain 3D wavenumber distributions of energy density ($D_{\epsilon_l}(\mathbf{k})$), where $\epsilon=V,B,N$ represents proton velocity ($V$), magnetic field ($B$), and proton density ($N$) fluctuations. In each bin, fluctuations have approximately the same wavenumber. To cover all MHD wavenumbers and ensure measurement reliability, we restrict our analysis to fluctuations with $1/(100d_{sc})<k<1.1\times0.1/d_i$ and $2/t^*<f_{rest}<f_{ci}/2$, where the wavenumber $k=\sqrt{k_\parallel^2+k_\perp^2}$, $t^*$ is the time window length, $f_{rest}=f_{sc}-\mathbf{k}\cdot \mathbf{V}/(2\pi)$ is the frequency in the plasma flow frame, and $\mathbf{V}$ is approaching the proton bulk velocity due to negligible spacecraft velocity. Fluctuations beyond the wavenumber and frequency ranges are set to zero. $D_{\epsilon_l}(\mathbf{k})$ is calculated by averaging $P_{\epsilon_l}(t,f_{sc})$ over effective time points in all time windows and integrating over $f_{sc}$. 

Seventh, the energy density of fast and slow modes is calculated by  $D_{\epsilon,\pm}=\delta\epsilon_{k,\pm}^2$, where '$+$' is for fast modes, '$-$' is for slow modes. The velocity fluctuations of fast and slow modes in wavevector space are calculated by 
\begin{eqnarray}
\delta V_{k,\pm} = \langle|\delta V_{k,\parallel}\hat{\mathbf{e}}_\parallel\cdot \hat{\mathbf{\xi}}_{\pm} \pm \delta V_{k,\perp2}\hat{\mathbf{e}}_{\perp2}\cdot \hat{\mathbf{\xi}}_{\pm}|\rangle
\label{eq:2},
\end{eqnarray}
where $\langle...\rangle$ denotes the average over effective time points in all time windows. In wavevector space, $\delta V_{k,\parallel}=\sqrt{\sum_{f_{sc}}D_{V_\parallel}(\mathbf{k},f_{sc},t)}\exp(2\pi i\phi_1(\mathbf{k},f_{sc},t))$, and $\delta V_{k,{\perp2}}=\sqrt{\sum_{f_{sc}}D_{V_{\perp2}}(\mathbf{k},f_{sc},t)}\exp(2\pi i\phi_2(\mathbf{k},f_{sc},t))$, where $\phi_1$ and $\phi_2$ are phase angles. The phase angle of turbulence fluctuations in a homogeneous system is usually assumed to be a uniform distribution. Therefore, we take $\phi_1$ and $\phi_2$ to be uniform in $[0,2 \pi]$, and Eq.(\ref{eq:2}) can be simplified to
\begin{eqnarray}
\delta V_{k,\pm}\propto |\sqrt{D_{V_\parallel}(\mathbf{k})}\cos\zeta_{\hat{e}_{\parallel}\hat{\mathbf{\xi}}_{\pm}}\pm \sqrt{D_{V_{\perp 2}}(\mathbf{k})}\sin\zeta_{\hat{e}_{\parallel}\hat{\mathbf{\xi}}_{\pm}}| \nonumber,
\end{eqnarray}
where $\zeta_{\hat{e}_{\parallel}\hat{\mathbf{\xi}}_{\pm}}$ is the angle between $\mathbf{\hat{e}}_{\parallel}$ and $\mathbf{\hat{\xi}}_{\pm}$. Fast- and slow-mode displacement vectors are given by \citep{Cho2003}
\begin{eqnarray}
\mathbf{\xi}_{\pm} \propto (-1+\alpha \pm \sqrt{A})k_{\parallel}\hat{\mathbf{e}}_{\parallel}+(1+\alpha \pm \sqrt{A})k_{\perp} \hat{\mathbf{e}}_{\perp 2}
\label{eq:3}.
\end{eqnarray}
The unit displacement vectors $\hat{\mathbf{\xi}}_{\pm}=
\mathbf{\xi}_{\pm}/
|\mathbf{\xi}_{\pm}|$. The parameter $A=(1+\alpha)^2-4\alpha \cos\theta$, where $\alpha=V_S^2/V_A^2$, $V_A$ is the Alfvén speed, $V_S$ is the sound speed, and $\theta$ is the angle between $\mathbf{k}$ and $\mathbf{B}_0$. 

Fast- and slow-mode magnetic field and proton density fluctuations are estimated as \citep{Cho2003} 
\begin{eqnarray}
\delta B_{k,\pm}=B_0\frac{\delta V_{k,\pm}}{V_{ph,\pm}}|\hat{\mathbf{e}}_{\parallel}\times\hat{\mathbf{\xi}}_{\pm}|,
\label{eq:4}\\
\delta N_{k,\pm}=N_0\frac{\delta V_{k,\pm}}{V_{ph,\pm}}\hat{\mathbf{k}}\cdot\hat{\mathbf{\xi}}_{\pm}
\label{eq:5}.
\end{eqnarray}
The unit wavevector $\hat{\mathbf{k}}=\mathbf{k}/|\mathbf{k}|$, and $N_0$ is the background proton density. Fast- and slow-mode phase speeds are given by \citep{Hollweg1975}
\begin{eqnarray}
V_{ph,\pm}^2=&&\frac{1}{2}\{(V_S^2 + V_A^2) \pm \nonumber\\ 
&&[(V_S^2 + V_A^2)^2 - 4V_S^2V_A^2cos^2\theta]^{1/2}\}.
\label{eq:6}
\end{eqnarray}
Appendix C shows that the decomposed magnetic field and density fluctuations (inferred from proton velocity fluctuations via Eqs.(\ref{eq:4},\ref{eq:5}) \citep{Cho2003}) match those directly measured by FGM and CIS-HIA instruments, indicating the reliability of MHD mode decomposition. All symbols used in this study are summarized in Table II (see Appendix).

\begin{figure}
\includegraphics[scale=0.145]{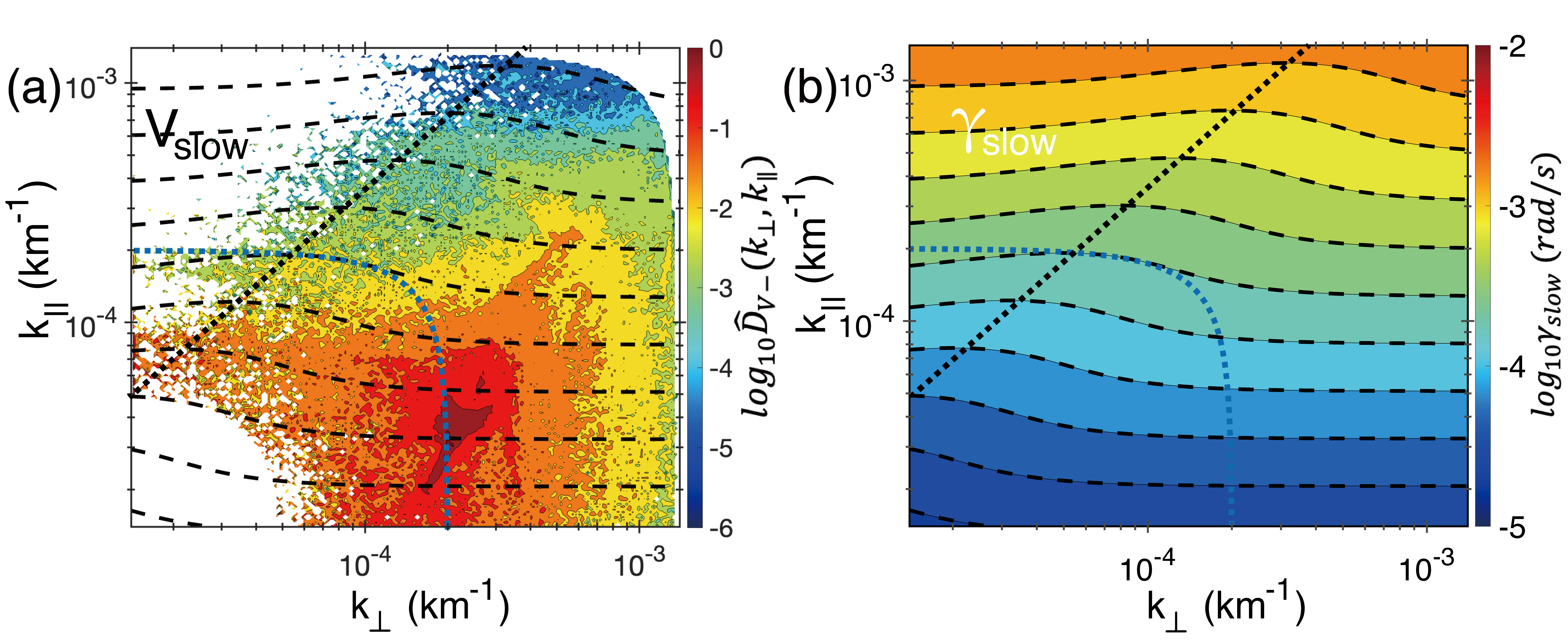}
\caption{Slow modes. (a) Wavenumber distributions of slow-mode proton velocity energy ($\hat{D}_{V-}$; color contours) and damping rate ($\gamma_{slow}$; black dashed curves). $\hat{D}_{V-}=D_{V-}/D_{V-,max}$ is normalized by the maximum energy density, where $\hat{D}_{V-}$ less than six orders of maximum magnitude or at $k<1/(100d_{sc})=5\times 10^{-5} km^{-1}$ is set to zero. (b) Wavenumber distributions of $\gamma_{slow}$. The black dotted line marks the peak line of $\gamma_{slow}$ contours. The blue dotted curve in each panel marks an isotropic contour at $k=2\times 10^{-4} km^{-1}$ as a reference.}
\end{figure}

\begin{figure*}
\includegraphics[scale=0.20]{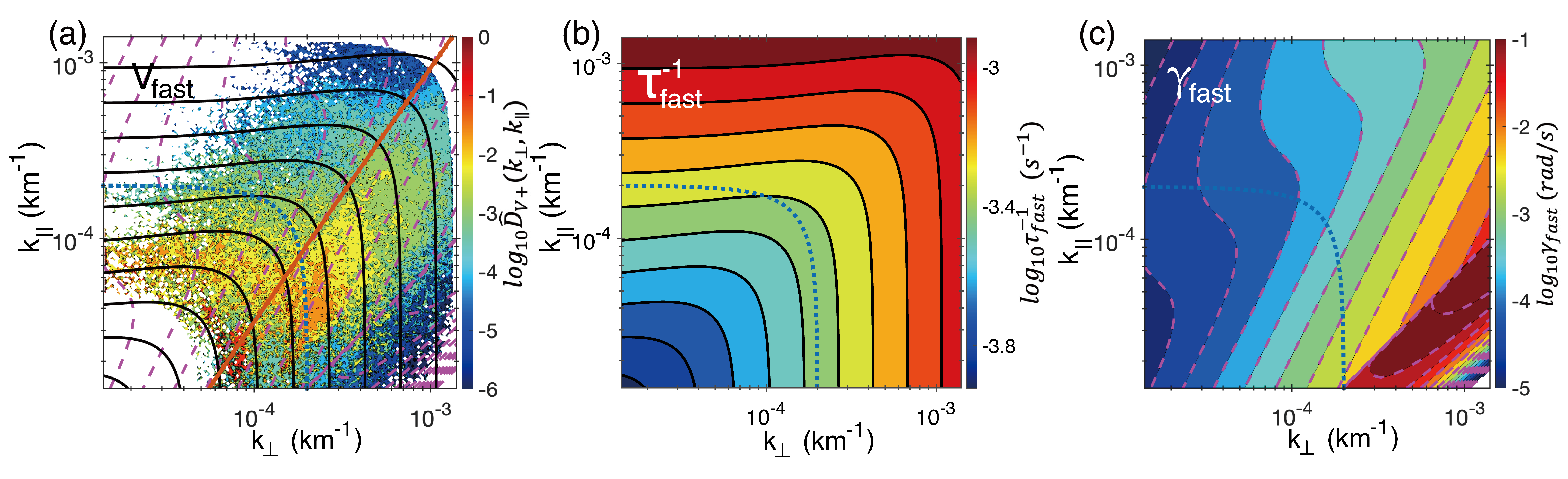}
\caption{Fast modes. (a) Wavenumber distributions of fast-mode proton velocity energy ($\hat{D}_{V+}$; color contours), cascading rate ($\tau_{fast}^{-1}$; black solid curves), and damping rate ($\gamma_{fast}$; purple dashed curves). To facilitate the comparison, the velocity energy spectrum is normalized by the same constant as slow modes in Figure 2(a). $\hat{D}_{V+}$ less than six orders of magnitude or at $k<1/(100d_{sc})=5\times 10^{-5} km^{-1}$ is set to zero. The red line marks the fast-mode CD truncation scales. (b) Wavenumber distributions of $\tau_{fast}^{-1}$. (c) Wavenumber distributions of $\gamma_{fast}$. The blue dotted curve in each panel marks an isotropic contour at $k=2\times 10^{-4} km^{-1}$ as a reference.}
\end{figure*}

\section{Slow modes} 

In Figure 2(a), we observe that the normalized wavenumber distributions of the proton velocity energy of slow modes ($\hat{D}_{V-}$; color contours) are prominently distributed along the $k_{\perp}$ axis, suggesting a faster cascade in the perpendicular direction. We also observe an increase in the anisotropy of energy distributions with the increasing wavenumbers. These observations indicate that smaller eddies of slow modes are more elongated along $\mathbf{B_0}$, in agreement with theoretical expectations and simulation results \citep{Cho2000,Makwana2020}. The anisotropic behaviors of slow modes found here are roughly similar to those for Alfvén modes \citep{Zhao2023}, presumably because slow modes passively mimic Alfvén modes \citep{Cho2003,Cho2005}.

The slow-mode theoretical damping rate ($\gamma_{slow}$) can be expressed as a function of wavenumber:
\begin{eqnarray}
\gamma_{slow}=&&\frac{|{\bf k}|V_S}{2|\cos \theta|}(\frac{1}{8}\pi\frac{m_e}{m_p})^{1/2}\nonumber \cdot\\
&&\left(1-\frac{\cos 2\theta[(V^2_S/V^2_A)\cos 2\theta-1]}{[1+V^4_S/V^4_A-2(V^2_S/V^2_A)\cos 2\theta]^{1/2}}\right)
\label{eq:7}
\end{eqnarray}
, where background parameters are listed in Table \ref{tab:table1}, and $m_e$ and $m_p$ are the electron and proton mass, respectively \citep{Oraevsky1983}. As shown in Figure 2(b) or Eq.(\ref{eq:7}), $\gamma_{slow}$ is very sensitive to the changes of $k_\parallel$ but not $k_\perp$. 
Interestingly enough, the constant contours of $\gamma_{slow}$ achieve peak values along the black dotted line in Figure 2(b). As it turns out, this special feature predicted by Eq.(\ref{eq:7}) is a notable signature of the CD modulation on energy spectra.

In Figure 2(a), $\gamma_{slow}$ contours (black dashed curves) are superposed on the $\hat{D}_{V-}$ spectrum (color contours). The similar pattern of energy iso-contours peaking along the black dotted line is also recognized in the upper left corner of the $\hat{D}_{V-}$ spectrum. This special consistency between the $\hat{D}_{V-}$ spectrum and $\gamma_{slow}$ contours is not seen in the Alfven mode counterpart \citep{Zhao2023}, where Alfven-mode energy steadily decreases with $k_\perp$ at each $k_\parallel$. Combined with the previous work \citep{Zhao2023}, our analysis in Figure 2 suggests that collisionless damping (CD) can {\it weakly} modulate energy distributions of slow modes but has little influence on those of Alfvén modes. Nevertheless, we note that this weak CD modification can be only observed in the upper left corner of the $\hat{D}_{V-}$ spectrum, which is likely because (1) $\gamma_{slow}$ increases with the increase of $k_\parallel$; (2) $\gamma_{slow}$ plays a limited role in shaping slow-mode energy spectra since the slow-mode cascading rate ($\tau_{slow}^{-1}$) is much larger than $\gamma_{slow}$ ($\tau_{slow}^{-1}/\gamma_{slow}\geq10$), where $\tau_{slow}^{-1}=k_\perp^{2/3}L_0^{-1/3}V_{A}$ \citep{Lithwick2001,Cho2003}. To simplify, the injection scale $L_0=T_c \delta V_0$, the correlation time $T_c$ determined by correlation functions is around $2300 s$ \citep{Zhao2023}, and the injection fluctuating velocity $\delta V_0$ is approximated as $M_{A,turb}V_A\sim 40km/s$.

To further estimate the anisotropy of fast and slow modes, we define a parameter
\begin{eqnarray}
R_{\pm}(k^{'})=\frac{D_{V\pm}(k_{\perp} =k^{'})}{D_{V\pm}(k_{\parallel} =k^{'})} 
\label{eq:8}
\end{eqnarray}
for a given wavenumber $k^{'}$, where the reduced perpendicular and parallel wavenumber distributions of the energy density are calculated by
\begin{eqnarray}
D_{\epsilon \pm}(k_{\perp})\sim \sum_{k_{\parallel}=k_{low}}^{k_{\parallel}=k_{upp}}D_{\epsilon \pm}(k_{\perp},k_{\parallel}) 
\label{eq:9},\\
D_{\epsilon \pm}(k_{\parallel})\sim \sum_{k_{\perp}=k_{low}}^{k_{\perp}=k_{upp}}D_{\epsilon \pm}(k_{\perp},k_{\parallel})
\label{eq:10}.
\end{eqnarray}
The integral upper limit $k_{upp}=0.1/d_i\sim1.4\times10^{-3} km^{-1}$, and integral lower limit $k_{low}=1/(100d_{sc})\sim5\times10^{-5} km^{-1}$.

Figure 4(a) shows a strong correlation between $log_{10}R_{-}(k^{'})$ and $log_{10}\gamma_{slow,\parallel}(k^{'})$ with the correlation coefficient of $0.97$, where the parallel damping rate $\gamma_{slow,\parallel}(k^{'})$ is calculated by averaging $\gamma_{slow}(k_\perp,k_\parallel=k^{'})$ over $k_\perp$ for a given wavenumber $k^{'}$. Moreover, the positive index of power-law fit ($0.46\pm0.28$) indicates that more significant anisotropy of slow modes corresponds to larger values of $\gamma_{slow}$ along $\mathbf{B}_0$.

\begin{figure}
\includegraphics[scale=0.16]{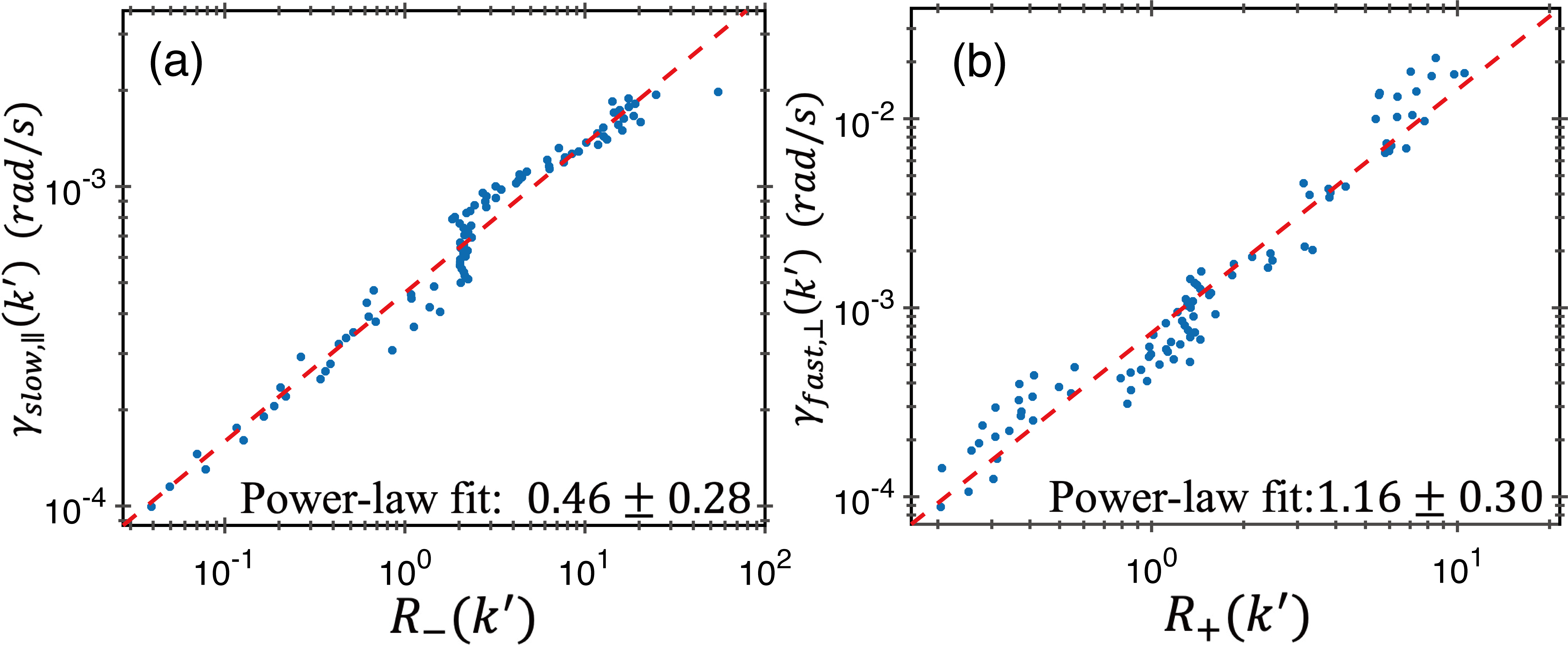}
\caption{The relations between the anisotropy of energy distributions ($R_{\pm}$ estimated by Eq.(\ref{eq:8})) and collisionless damping rates ($\gamma_{slow}$ and $\gamma_{fast}$ estimated by Eqs.(\ref{eq:7},\ref{eq:11},\ref{eq:12})). (a) $R_{-}$ versus $\gamma_{slow,\parallel}$. (b) $R_{+}$ versus $\gamma_{fast,\perp}$. The red dashed lines represent power-law fits. The shown average values and standard deviations of power-law indices are obtained using a least-squares fit to the relations on a log–log plot.}
\end{figure}

\section{Fast modes}

In Figure 3(a), we observe that the normalized wavenumber distributions of the proton velocity of fast modes ($\hat{D}_{V+}$; color contours) show more isotropic features than slow modes. Figure 3(b) shows that wavenumber distributions of the fast-mode cascading rate ($\tau_{fast}^{-1}$) present an scale-independent anisotropy, where $\tau_{fast}^{-1}\sim(k/L_0)^{1/2}\delta V_0^2/V_{ph,+}$ \citep{Yan2004}. When $\omega_{fast}/k_\parallel\ll V_{Te}$, the fast-mode theoretical damping rate is given by \citep{Oraevsky1983}:
\begin{eqnarray}
\gamma_{fast}=&&\frac{|{\bf k}|V_S}{2|\cos \theta|}(\frac{1}{8}\pi\frac{m_e}{m_p})^{1/2}\nonumber \cdot\\
&&\left(1+\frac{\cos 2\theta[(V^2_S/V^2_A)\cos 2\theta-1]}{[1+V^4_S/V^4_A-2(V^2_S/V^2_A)\cos 2\theta]^{1/2}}\right),
\label{eq:11}
\end{eqnarray}

When $\omega_{fast}/k_\parallel\gg V_{Te}$, 
\begin{eqnarray}
\gamma_{fast}=&&(\frac{1}{8}\pi)^{1/2}\omega_{fast}\frac{m_eV_{Te}sin^2\theta}{m_pV_A|cos\theta|}exp(-\frac{\omega_{fast}^2}{2k_\parallel^2V^2_{Te}}),
\label{eq:12}
\end{eqnarray}
$\omega_{fast}$ is the fast-mode frequency, $V_{Te}=\sqrt{k_BTe/m_e}$ is the electron thermal speed, and $k_B$ is the Boltzmann constant. Figure 3(c) shows $\gamma_{fast}$ calculated by combining Eq.(\ref{eq:11}) at $\omega_{fast}/k_\parallel\leq V_{Te}$ and Eq.(\ref{eq:12}) at $\omega_{fast}/k_\parallel\geq V_{Te}$, where the background parameters are listed in Table 1. With increasing $k_{\perp}$, $\gamma_{fast}$ sharply enhances and is up to $0.9 rad/s$ in the bottom right corner where $\theta$ approaches $90^\circ$, indicating that fast modes undergo more severe CD damping and rapid dissipation at larger $k_{\perp}$ and larger $\theta$. 

We superpose the contours of $\tau_{fast}^{-1}$ (black solid curves) and $\gamma_{fast}$ (purple dashed curves) on the $\hat{D}_{V+}$ spectrum in Figure 3(a). Different from slow modes, fast modes exhibit CD truncation scales ($kc$; the red line) which are obtained by equating $\tau_{fast}^{-1}$ and $\gamma_{fast}$ \citep{Yan2008}. On $kc$ scales, turbulence cascade and CD play a comparable role in shaping energy distributions. (1) At $k<2\times 10^{-4} km^{-1}$ (blue dotted curve), the $\hat{D}_{V+}$ spectrum agrees well with the $\tau_{fast}^{-1}$ contours on the left side of the red line (above $kc$; $\tau_{fast}^{-1}>\gamma_{fast}$). The weak, scale-independent anisotropy of the $\hat{D}_{V+}$ spectrum suggests that fast-mode energy distributions do not depend heavily on $\mathbf{B}_0$, consistent with numerical simulations and theoretical analysis of fast modes \citep{Cho2003,Galtier2023}. (2) At $k>2\times 10^{-4} km^{-1}$, with the increasing $k_\perp$, the magnitude of the $\hat{D}_{V+}$ spectrum generally decreases, whereas $\gamma_{fast}$ sharply increases on the right side of the red line (below $kc$; $\tau_{fast}^{-1}<\gamma_{fast}$). These results imply that $\gamma_{fast}$ likely plays a more crucial role in shaping the fast-mode energy spectrum on smaller scales. Above all, these findings support the theoretical prediction that energy distributions of fast-mode turbulence are shaped by the forcing on large scales and the damping on small scales.

Figure 4(b) shows that the correlation coefficient between $log_{10}R_{+}(k^{'})$ and $log_{10}\gamma_{fast,\perp}(k^{'})$ is 0.97, and their power-law fit is $1.16\pm0.30$, where the perpendicular damping rate $\gamma_{fast,\perp}(k^{'})$ is obtained by averaging $\gamma_{fast}(k_\perp=k^{'},k_\parallel)$ over $k_\parallel$ for a given wavenumber $k'$. This strong correlation between CD and anisotropy of fast modes further indicates that CD plays an increasingly important role in shaping energy distributions of fast modes as $k_\perp$ increases.

\begin{figure}
\includegraphics[scale=0.155]{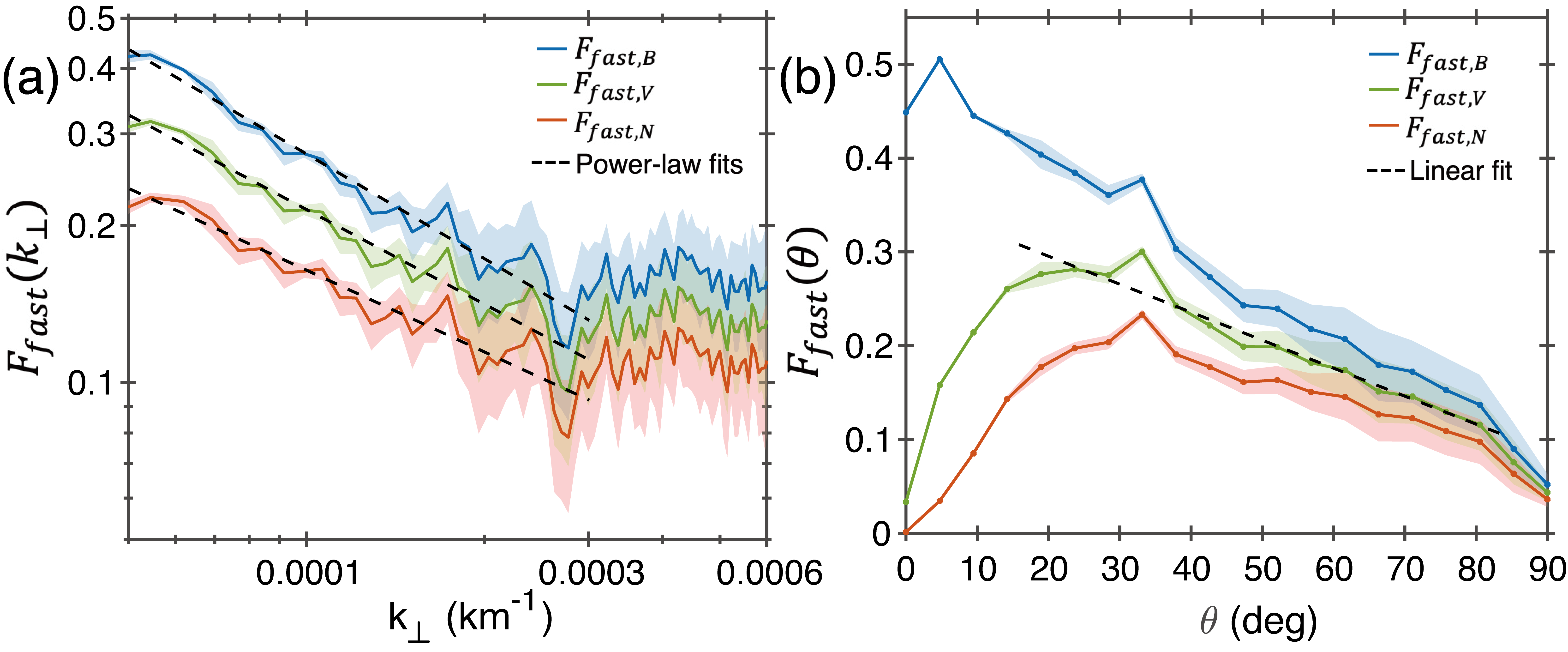}
\caption{The fast mode fraction $F_{fast,\epsilon}=\frac{D_{\epsilon +}}{D_{\epsilon +}+D_{\epsilon -}}$ calculated with $k_{low}< k_\perp,k_\parallel<k_{upp} $(see Eqs.(\ref{eq:9},\ref{eq:10})). (a) $F_{fast,\epsilon}(k_\perp)$ versus $k_\perp$. (b) $F_{fast,\epsilon}(\theta)$ versus $\theta$. Blue, green, and red curves represent energy fractions of magnetic field ($B$), proton velocity ($V$), and proton density ($N$), respectively. The color-shaded regions represent standard deviations of $F_{fast,\epsilon}$ using data sets from $\eta<10^\circ$ to $\eta<30^\circ$ (binning $\Delta \eta=1^\circ$). The black dashed curves represent the power-law fits of $F_{fast,\epsilon}(k_\perp)$ or linear fit of $F_{fast,V}(\theta)$.}
\end{figure}

\section{Fast mode fraction relative to the total compressional modes} 

Figure 5 shows the ratio of fast-mode energy to the total energy of compressible modes as a function of $k_\perp$ and $\theta$, which is defined as $F_{fast,\epsilon}=\frac{D_{\epsilon +}}{D_{\epsilon +}+D_{\epsilon -}}$, where $\epsilon=V,B,N$ represents proton velocity ($V$), magnetic field ($B$), and proton density ($N$) fluctuations. For easier reference, we name this ratio 'fast mode fraction'. The color-shaded regions in both panels of Figure 5 represent standard deviations of the results for data sets from $\eta<10^\circ$ to $\eta<30^\circ$, binning $\Delta \eta = 1^\circ$. It is evident from Figure 5 that the $\eta$ deviation does not affect the general trend of $F_{fast,\epsilon}$ as functions of $k_\perp$ and $\theta$. In Figure 5(a), $F_{fast,\epsilon}$ has a decreasing trend as $k_\perp$ increases up to $k_\perp<3\times10^{-4} km^{-1}$, after which $F_{fast,\epsilon}$ is roughly a constant with a small value $[0.1,0.2]$. Furthermore, regression analysis suggests that the three fractions $F_{fast,\epsilon}$ have similar power-law scaling at $k_\perp < 3\times10^{-4} km^{-1}$, indicating that CD damps out the fast mode fraction consistently over all MHD observables.

We perform a similar analysis on $F_{fast,\epsilon}$ to $\theta$ to explore whether there is an angle dependence for CD modulation. Figure 5(b) demonstrates that, at small $\theta$, fast modes dominate magnetic field fluctuations, whereas slow modes dominate proton density fluctuations. As $\theta$ increases, fast mode fractions of the three parameters gradually become comparable. Moreover, the linear decrease of $F_{fast,V}$ as $\theta$ increases is consistent with higher $\gamma_{fast}$ at larger $\theta$. The different $\theta$ dependencies in $F_{fast,B}$ and $F_{fast,N}$ can be attributed to the angle dependence in the calculations of Eqs.(\ref{eq:4},\ref{eq:5}), wherein magnetic field and proton density fluctuations are inferred from proton velocity fluctuations.

\section{Summary}
This study presents the first observational evidence of substantial CD modulation on compressible MHD turbulence cascade. Utilizing an improved MHD mode decomposition technique, we are able to obtain wavenumber distributions of slow and fast modes via four \textit{Cluster} spacecraft measurements in Earth's magnetosheath. Our findings are summarized below:

(1) Wavenumber distributions of slow modes are mainly stretched perpendicular to $\mathbf{B_0}$ and weakly modulated by CD. In contrast, fast modes are subject to a more significant CD modulation. Fast modes exhibit a weak, scale-independent anisotropy above the CD truncation scales. Below the CD truncation scales, the anisotropy of fast modes enhances as wavenumbers increase. These observations provide the first observational evidence for damping affecting the small-amplitude compressible MHD turbulence cascade on smaller scales within the MHD regime.


(2) Due to the strong dependence on wave propagation angle, CD increases the slow-mode anisotropy parallel to $\mathbf{B}_0$, whereas CD increases the fast-mode anisotropy perpendicular to $\mathbf{B}_0$.

(3) Fast mode fractions in the total energy of compressible modes are scale- and angle-dependent, which decrease as $k_\perp$ (or $\theta$) increases.

These observational results are consistent with theoretical expectations \citep{Yan2004,Yan2008}. Because the plasma parameters in the analyzed event are common in astrophysical and space plasma systems, our results improve the general understanding of the role of CD in the cascade of compressible turbulence and the corresponding energy transfer, particle transport, and particle energization.

\section{Acknowledgments}
We would like to thank the members of the \textit{Cluster} spacecraft team and NASA’s Coordinated Data Analysis Web. The \textit{Cluster} data are available at \url{https://cdaweb.gsfc.nasa.gov}. Data analysis was performed using the IRFU-MATLAB analysis package available at \url{https://github.com/irfu/irfu-matlab}. K.H.Y acknowledges the support from the Laboratory Directed Research and Development program of Los Alamos National Laboratory under project number(s) 20220700PRD1.

%




\appendix

\section{Details of examination of the turbulence state}

To examine the turbulent state, we calculate the normalized correlation function $R(\tau)/R(0)$, where the correlation function is defined as $R(\tau)=\langle\delta B(t)\delta B(t+\tau)\rangle$, $\tau$ is the timescale, and angular brackets represent a time average over the time window length (5 hours). Figure 7 shows $R(\tau)/R(0)$ for magnetic field $\delta B_{\perp 1}$ and $\delta B_{\perp 2}$ components in field-aligned coordinates. Fluctuations $\delta B_{\perp 1}$ are in $(\hat{\mathbf{b}}_0\times\hat{\mathbf{X}}_{GSE})\times\hat{\mathbf{b}}_0$ directions, and $\delta B_{\perp 2}$ are in $\hat{\mathbf{b}}_0\times\hat{\mathbf{X}}_{GSE}$ directions, where $\hat{\mathbf{X}}_{GSE}$ is the unit vector towards the Sun from the Earth.  

The correlation time is estimated as $T_c \sim \int ^{R(\tau)\rightarrow\frac{1}{2e}}_0 R(\tau)/R(0)d\tau\sim[1300,2300]s$. Thus, $T_c$ is much less than the time window length (5 hours), suggesting that fluctuations are approximately stationary. Moreover, $R(\tau)/R(0)$ profiles in all time windows are similar, suggesting that the starting time of the moving time window has a slight influence on $R(\tau)/R(0)$, and thus fluctuations are homogeneous. Above all, it is reasonable to describe structures of turbulent fluctuations using three-dimensional energy distributions.

\section{Examination of the effects of angle $\eta$}

Figure 8 shows wavenumber distributions of proton velocity energy density from three representative data sets ($\eta<10^\circ$, $\eta<20^\circ$, and $\eta<30^\circ$). In general, the main properties of energy spectra of slow and fast modes do not significantly change with the increase of $\eta$. We note that the anisotropic signature of fast modes is more prominent when we relax the constraints on $\eta$. It may be because the mode decomposition between slow and fast modes becomes more incomplete when using a more relaxed $\eta$ constraint. The $\eta$ effects on energy distributions are also shown in the shaded regions of Figure 5.  

\section{Examination of MHD mode decomposition}

To examine the reliability of MHD mode decomposition, we compare decomposed magnetic field and density fluctuations (inferred from proton velocity fluctuations using Eqs.(\ref{eq:4},\ref{eq:5})) with those directly measured by FGM and CIS-HIA instruments. According to the linearized induction equation, the magnetic field within $\hat{k}\hat{b}_0$ plane fluctuates along the wave front $\mathbf{e}_{wf}=\hat{\mathbf{k}}\times (\hat{\mathbf{e}}_\parallel\times\hat{\mathbf{k}})$ (the yellow dashed line in Figure 6(b)). The strength of magnetic field fluctuations within the $\hat{k}\hat{b}_0$ plane is directly measured by FGM instruments.
\begin{eqnarray}
    |\delta \mathbf{B}_{obs,in plane}|=\langle|\delta B_{k,\parallel}\hat{\mathbf{e}}_\parallel\cdot\mathbf{e}_{wf} + 
    \delta B_{k,{\perp 2}}\hat{\mathbf{e}}_{\perp 2}\cdot\mathbf{e}_{wf}|\rangle.
\label{eq:13}
\end{eqnarray}
In wavevector space,
\begin{eqnarray}
\delta B_{k,\parallel}=\sqrt{\sum_{f_{sc}}D_{B_\parallel}(\mathbf{k},f_{sc},t)}exp(2\pi i\phi_1(\mathbf{k},f_{sc},t)),\\
\delta B_{k,{\perp2}}=\sqrt{\sum_{f_{sc}}D_{B_{\perp2}}(\mathbf{k},f_{sc},t)}exp(2\pi i\phi_2(\mathbf{k},f_{sc},t)).
\end{eqnarray}
Similar to Eq.(\ref{eq:2}), we take $\phi_1$ and $\phi_2$ to be uniform in $[0,2\pi]$, Eq.(\ref{eq:13}) would be simplified to 
\begin{eqnarray}
|\delta \mathbf{B}_{obs,inplane}|
\propto|\sqrt{D_{B_\parallel}(\mathbf{k})}\hat{\mathbf{e}}_\parallel\cdot\mathbf{e}_{wf} + \sqrt{D_{B_{\perp2}}(\mathbf{k})} \hat{\mathbf{e}}_{\perp 2}\cdot\mathbf{e}_{wf}|.
\end{eqnarray}
The directly observed magnetic field energy within $\hat{k}\hat{b}_0$ plane is calculated by $D_{B,obs,inplane}=|\delta \mathbf{B}_{obs,inplane}|^2$. It is worth noting that we have filtered out Alfven-mode fluctuations (fluctuations out of the $\hat{k}\hat{b}_0$ plane) in this process. On the other hand, magnetic field fluctuations within the $\hat{k}\hat{b}_0$ plane are composed of slow and fast modes, which are inferred from proton velocity fluctuations using Eq.(\ref{eq:4}) \citep{Cho2003}. Figures 9(a,b) show wavenumber distributions of slow- and fast-mode magnetic field energy ($D_{B-}$ and $D_{B+}$). In Figures 10(a,b), the total magnetic field energy of slow and fast modes ($D_{B,sum}=D_{B+}+D_{B-}$) are roughly consistent with $D_{B,obs,inplane}$. 

Different from magnetic field fluctuations, the proton density fluctuates along wavevectors; thus, all density fluctuations are within the $\hat{k}\hat{b}_0$ plane. The proton density energy directly observed by CIS-HIA instruments is given by $D_{N,obs}=\delta N_{obs}^2 \propto \delta N_{k}^2$. Moreover, based on the continuity equation (Eq.(\ref{eq:5})), slow and fast modes provide proton density fluctuations together, shown in Figures 9(c,d). In Figures 10(c,d), the total proton density energy of slow and fast modes ($D_{N,sum}=D_{N+}+D_{N-}$) agree well with $D_{N,obs}$. 

Figures 11(a,b) show 2D spectral ratios $\frac{\hat{D}_{B,sum}}{\hat{D}_{B,obs,inplane}}$ and $\frac{\hat{D}_{N,sum}}{\hat{D}_{N,obs}}$, respectively. Most ratios are around unity, suggesting the results of MHD mode decomposition are reliable. Moreover, 2D spectral ratios calculated by data sets from $\eta<10^\circ$ to $\eta<30^\circ$ (binning $\Delta \eta=1^\circ$) show similar results (not shown). 

\begin{figure}[ht]
\includegraphics[scale=0.3]{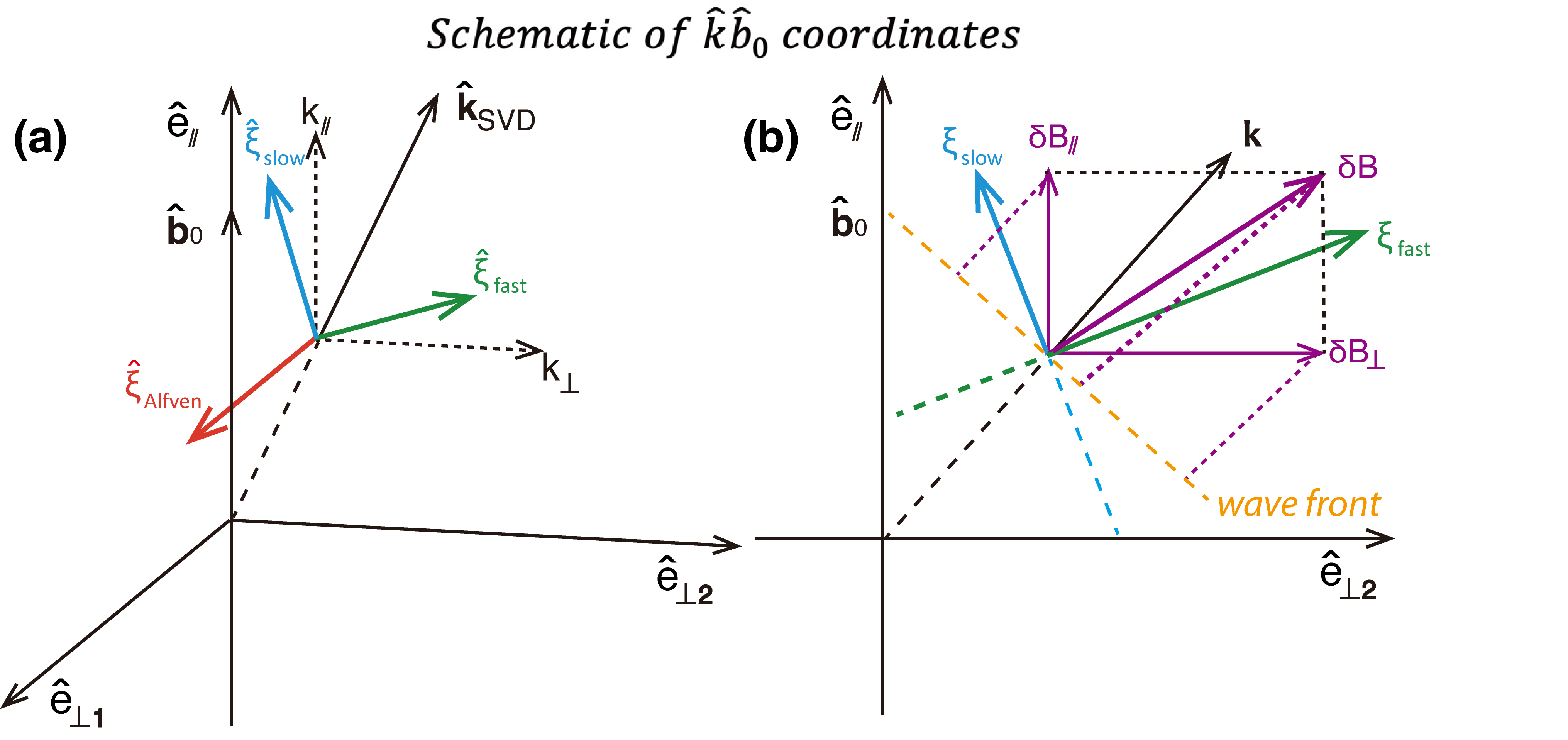}
\centering
\caption{(a) Schematic of $\hat{k}\hat{b}_0$ coordinates determined by $\hat{\mathbf{k}}_{SVD}$ and $\hat{\mathbf{b}}_0$. The red, blue, and green arrows represent the unit displacement vectors of Alfvén, slow, and fast modes ($\hat{\mathbf{\xi}}_{Alfven}$, $\hat{\mathbf{\xi}}_{slow}$, and $\hat{\mathbf{\xi}}_{fast}$). (b) The composition of magnetic field fluctuations (purple arrows). The yellow dashed line marks the wave front.}
\label{Figure 6}
\end{figure}

\begin{figure}[ht]
    \centering
    \includegraphics[scale=0.5]{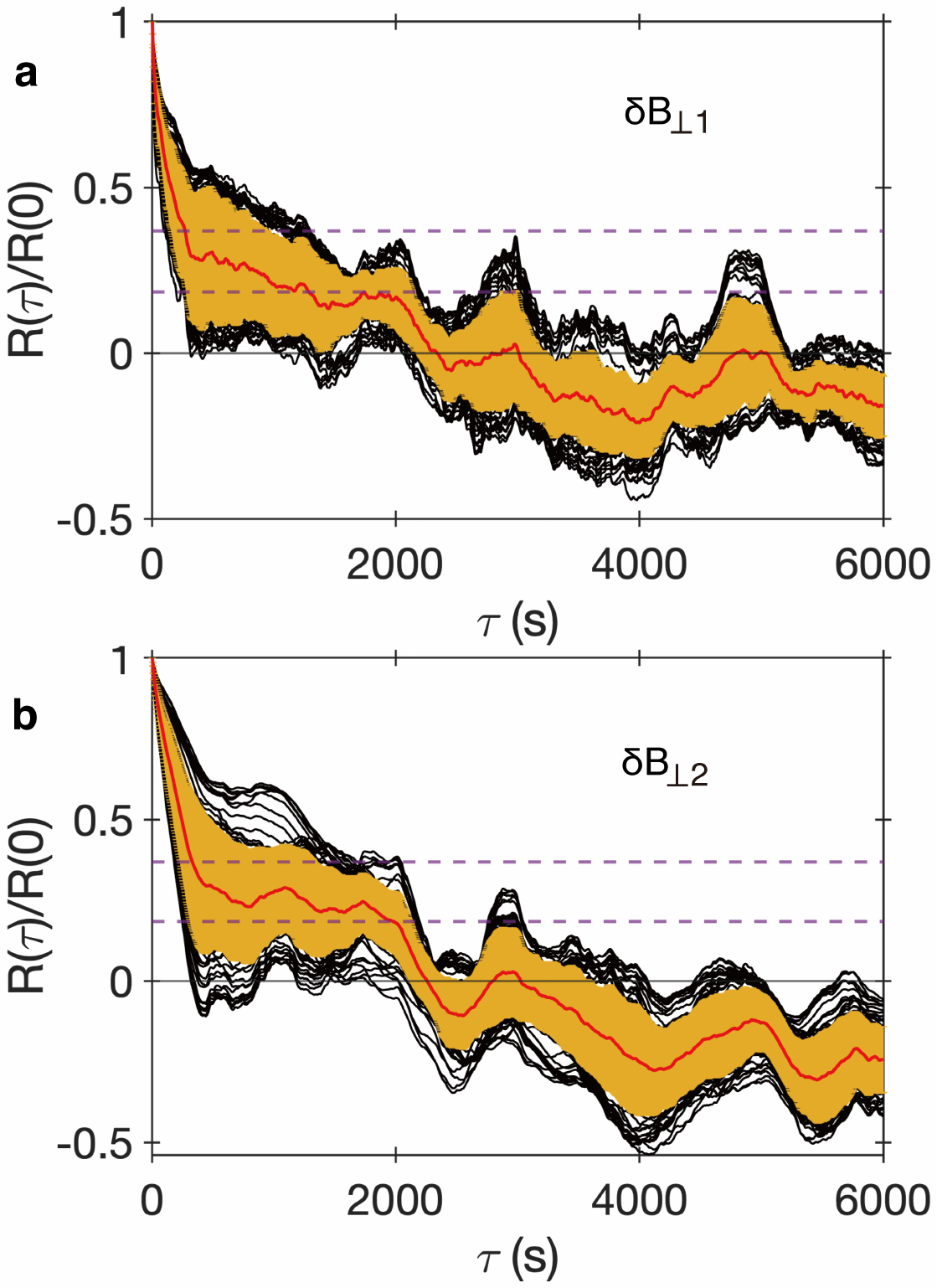}
    \caption{Normalized correlation functions $R(\tau)/R(0)$ versus $\tau$ for $\delta B_{\perp 1}$ and $\delta B_{\perp 2}$ in field-aligned coordinates. The black curves represent $R(\tau)/R(0)$ in all time windows. The red curves represent average values ($s$) of $R(\tau)/R(0)$ over all time windows. The yellow-shaded regions represent $[s-\sigma,s+\sigma]$, where $\sigma$ represents standard deviations of $R(\tau)/R(0)$. The purple horizontal dashed lines represent $R(\tau)/R(0)=1/e$ and $1/(2e)$.}
    \label{Figure 7}
 \end{figure}

\begin{figure}[ht]
\includegraphics[scale=0.2]{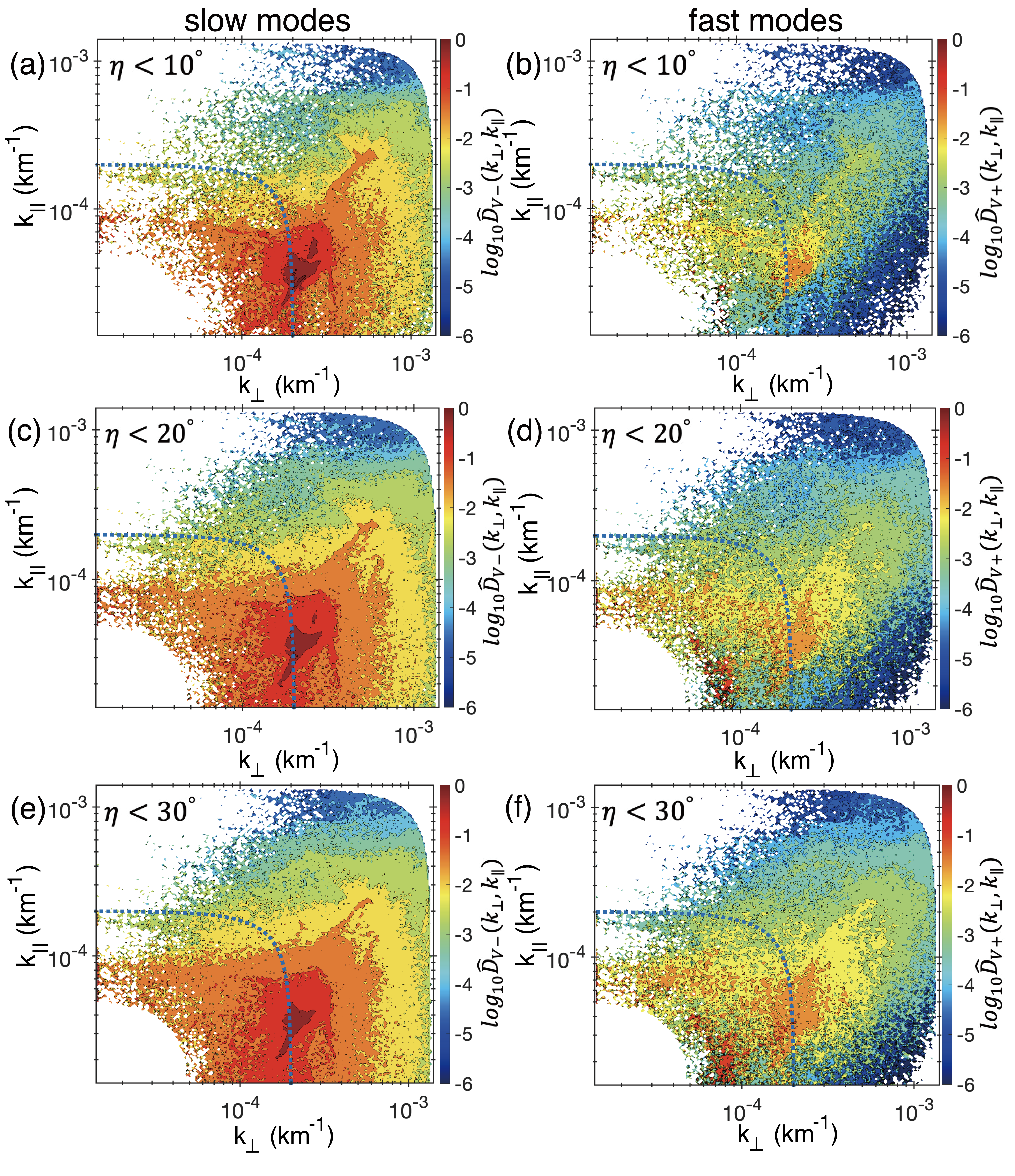}
\centering
\caption{Wavenumber distributions of proton velocity fluctuations under $\eta<10^\circ$ (a,b), $\eta<20^\circ$ (c,d), and $\eta<30^\circ$ (e,f). (a,c,e) slow-mode spectra. (b,d,f) fast-mode spectra. All spectra are normalized with the maximum energy density to facilitate comparison, where $\hat{D}_{\epsilon\pm}$ less than six orders of magnitude or at $k<1/(100d_{sc})=5\times 10^{-5} km^{-1}$ is set to zero. The blue dotted curve in each panel marks an isotropic contour at $k=2\times 10^{-4} km^{-1}$ as a reference.} 
\label{Figure 8}
\end{figure}

\begin{figure}[ht]
 \includegraphics[scale=0.3]{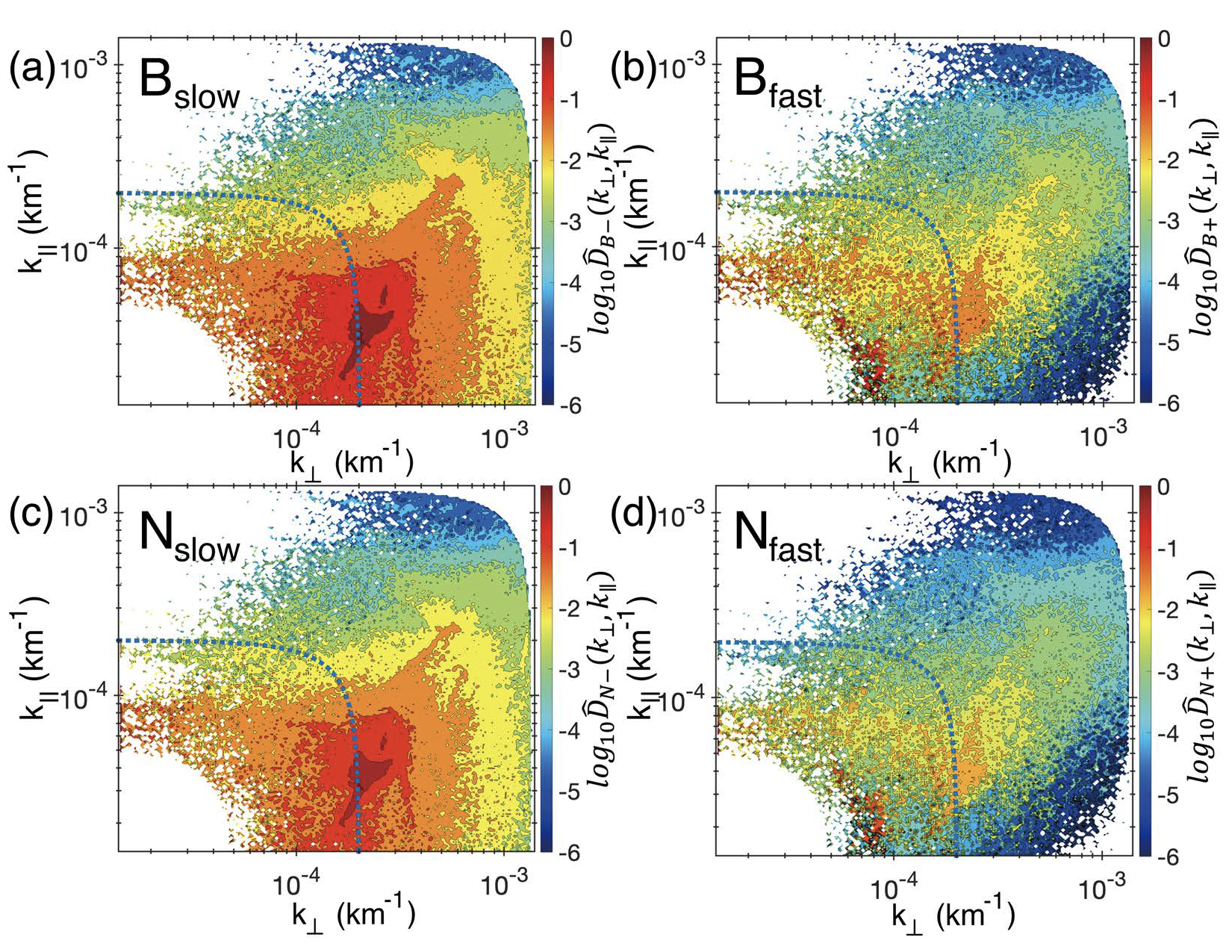}
 \centering
 \caption{(a,b) Wavenumber distributions of slow- and fast-mode magnetic field energy ($\hat{D}_{B-}$ and $\hat{D}_{B+}$), which are normalized with the same constant. (c,d) Wavenumber distributions of slow- and fast-mode proton density energy ($\hat{D}_{N-}$ and $\hat{D}_{N+}$), which are normalized with the same constant. $\hat{D}_{\epsilon\pm}$ less than six orders of magnitude or at $k<1/(100d_{sc})=5\times 10^{-5} km^{-1}$ is set to zero. The blue dotted curve in each panel marks an isotropic contour at $k=2\times 10^{-4} km^{-1}$ as a reference. These figures use data under $\eta<20^\circ$.}
 \label{Figure 9}
\end{figure}

\begin{figure}[ht]
\includegraphics[scale=0.3]{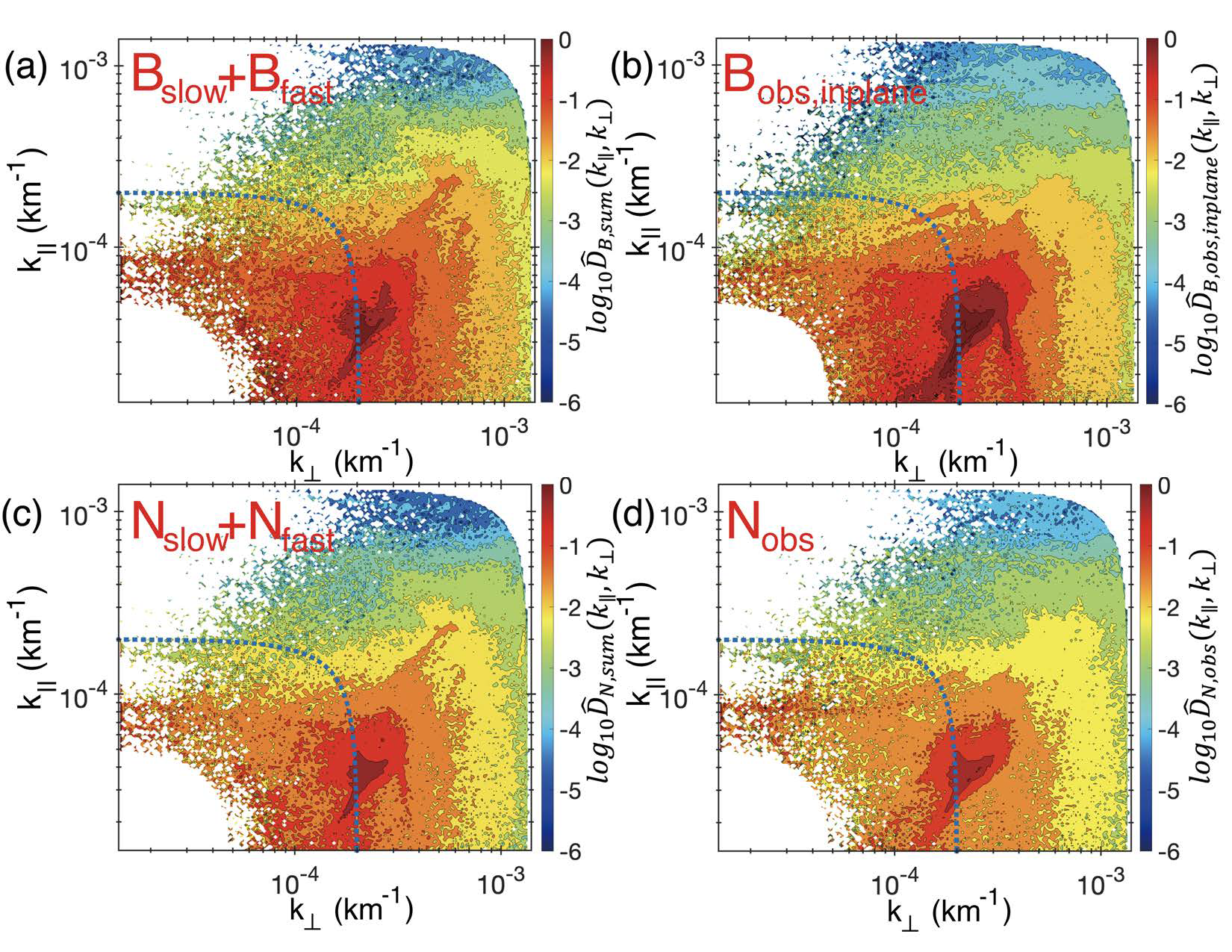}
\centering
\caption{Comparisons of decomposed and directly measured results. (a,c) Wavenumber distributions of magnetic field energy within $\hat{k}\hat{b}_0$ plane ($\hat{D}_{B,sum}$) and proton density energy ($\hat{D}_{N,sum}$) inferred from proton velocity energy. (b,d) Directly measured wavenumber distributions of magnetic field energy within $\hat{k}\hat{b}_0$ plane ($\hat{D}_{B,obs,inplane}$) and proton density energy ($\hat{D}_{N,obs}$). To reduce the effects of residual energy, spectrum (b) is normalized $m$ times the constant compared to (a), where $m=\langle M_{A,turb}/(\delta B/B_0)\rangle^2$. Spectra (c,d) are normalized with the same constant. The energy density less than six orders of magnitude or at $k<1/(100d_{sc})=5\times 10^{-5} km^{-1}$ is set to zero. The blue dotted curve in each panel marks an isotropic contour at $k=2\times 10^{-4} km^{-1}$ as a reference. These figures use data sets under $\eta<20^\circ$.}
\label{Figure 10}
\end{figure}


\begin{figure*}[ht]
\includegraphics[scale=0.2]{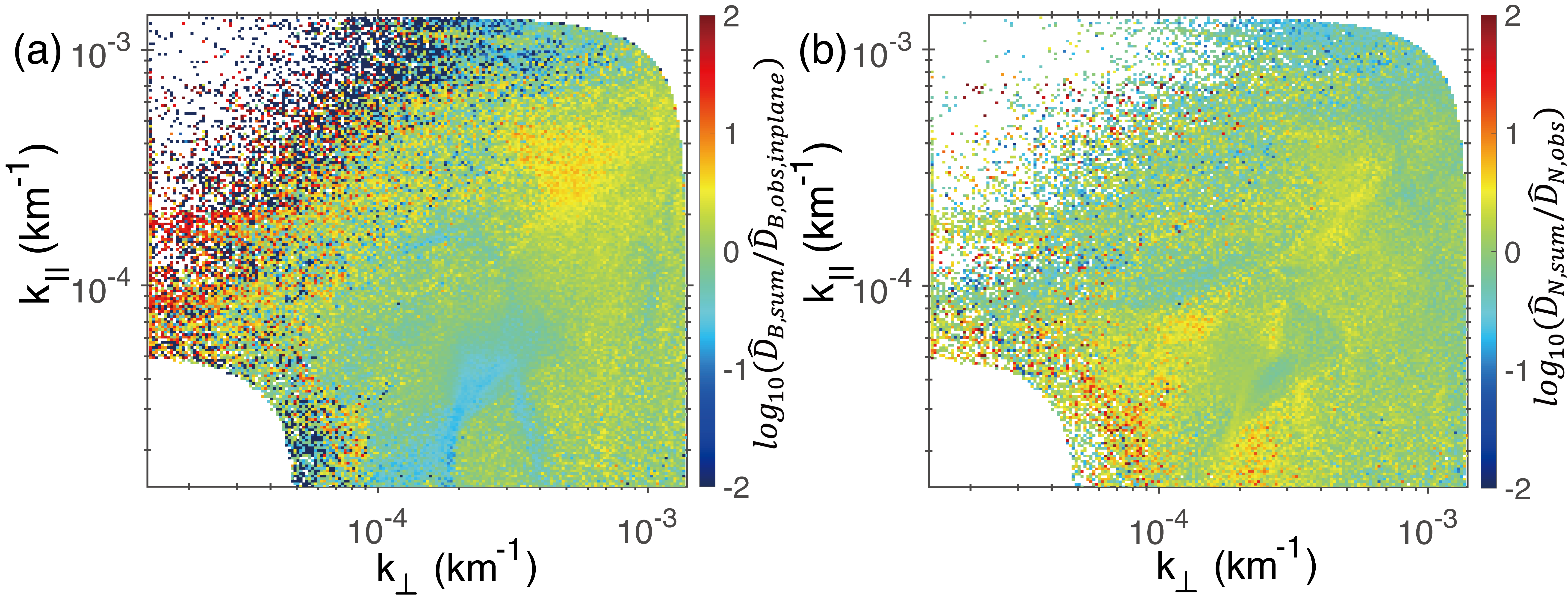}
\centering
\caption{Spectral ratios of decomposed and directly measured results. (a) $\frac{\hat{D}_{B,sum}}{\hat{D}_{B,obs,inplane}}$ (the ratio of Figure 10(a) to Figure 10 (b)). (b) $\frac{\hat{D}_{N,sum}}{\hat{D}_{N,obs}}$ (the ratio of Figure 10(c) to Figure 10 (d)). The ratios at $k<1/(100d_{sc})=5\times 10^{-5} km^{-1}$ are set to zero. These figures use data sets under $\eta<20^\circ$.}
\label{Figure 11}
\end{figure*}

\begin{table*}
\caption{\label{tab:table2} The definition of symbols.}
\begin{ruledtabular}
\begin{tabular}{cc}
 Symbol & Definition \\ \hline
 $\mathbf{B}_0$ & Background magnetic field \\
 $\delta B $  & RMS magnetic field fluctuations \\
 $\delta B_{k,\pm} $  & Magnetic field fluctuations of fast (slow) modes \\
 $\delta B_{obs,inplane} $  & Magnetic field fluctuations within $\hat{k}\hat{b}_0$ plane observed by FGM\\
 $N_0$ & Background proton density\\
 $\delta N_{k,\pm} $  & Proton density fluctuations of fast (slow) modes \\
$\delta N_{obs} $  & Proton density fluctuations of observed by CIS-HIA\\
 $V_A$ & Alfvén velocity\\
 $V_S$ & Sound velocity\\
 $V_{Te}$ & Electron thermal speed\\
 $V_{ph,\pm} $ & Phase velocity of fast (slow) modes\\
 $\delta V $  & RMS proton velocity fluctuations \\
 $\delta V_{k,\pm} $  & Proton velocity fluctuations of fast (slow) modes \\
 $M_{A,turb}$& Turbulent Alfvén Mach number \\
 $\beta_p$ & Ratio of proton thermal to magnetic pressure \\
 $d_{sc}$ & Spacecraft relative separation \\
 $d_{i}$  & Proton inertial length \\
 $k_\perp$ & Wavenumber perpendicular to background magnetic field  \\
 $k_\parallel $ &Wavenumber parallel to background magnetic field  \\
 $\mathbf{\hat{k}}_{SVD} $ & Wavenumber direction determined by singular value decomposition \\
 $\mathbf{k}$ ($\mathbf{k}_{B_l}$) & Wavevector determined by multi-spacecraft timing analysis \\
 $\mathbf{\xi}_{\pm}$  & Fast- and slow-mode displacement vector  \\
 $f_{sc}$ & Frequency in the spacecraft frame \\ 
 $f_{rest}$& Frequency in the plasma flow frame \\
 $\gamma_{fast}$, $\gamma_{slow}$& Fast- and slow-mode damping rate \\
 $\tau_{fast}$, $\tau_{slow}$& Fast- and slow-mode cascading rate \\
 $R_{\pm}$ & Anisotropy of fast and slow modes  \\
 $\theta$ & Angle between wavevector and background magnetic field \\
 $\eta$ & Angle between wavevector and $\hat{k}\hat{b}_0$ plane \\
 $\zeta$ & Angle between background magnetic field and displacement vector \\
 $\phi$ & Phase angle of fluctuations in Fourier space \\
 $W_{\epsilon_l}$ ($\epsilon\equiv V,B,N$) & Wavelet coefficients of fluctuations \\
 $P_{\epsilon_l}$ & Power of fluctuations\\
 $D_{\epsilon_l}$& Energy density of fluctuations\\
 $D_{B,obs,inplane}$, $D_{N,obs}$ & Directly measured energy density of in-plane magnetic field and proton density\\
$F_{fast,\epsilon}$ & Fast-mode energy to total energy of compressible modes\\
\end{tabular}
\end{ruledtabular}
\end{table*}

\bibliography{sample631}{}
\bibliographystyle{aasjournal}



\end{document}